\documentclass[%
 reprint,
superscriptaddress,
 amsmath,amssymb,
 aps,
floatfix,
]{revtex4-1}

\usepackage{comment}
\usepackage{hyperref}
\usepackage{cases}
\usepackage{graphicx}
\usepackage{dcolumn}
\usepackage{bm}
\usepackage{algorithm}
\usepackage{algorithmic}
\makeatletter
\newcommand{\HEADER}[1]{\ALC@it\underline{\textsc{#1}}\begin{ALC@g}}
\newcommand{\ENDHEADER}{\end{ALC@g}}
\makeatother

\begin{document}


\title{Perceptrons from Memristors}

\author{Francisco Silva}
\email{francisco.horta.ferreira.da.silva@tecnico.ulisboa.pt}
\affiliation{Instituto de Telecomunica\c{c}\~oes, Physics of Information and Quantum Technologies Group, Portugal}
\author{Mikel Sanz}%
\email{mikel.sanz@ehu.eus}
\affiliation{Department of Physical Chemistry, University of the Basque Country UPV/EHU, Apartado 644, E-48080 Bilbao, Spain}
\author{Jo\~ao Seixas}
\email{joao.seixas@tecnico.ulisboa.pt}
\affiliation{Instituto Superior T\'ecnico, Universidade de Lisboa, Portugal}
\affiliation{CeFEMA, Instituto Superior T\'ecnico, Universidade de Lisboa, Portugal}
\affiliation{Laborat\'orio de Instrumenta\c{c}\~ao e F\'isica Experimental de Part\'iculas (LIP), Lisbon, Portugal}
\author{Enrique Solano}
\email{enr.solano@gmail.com}
\affiliation{Department of Physical Chemistry, University of the Basque Country UPV/EHU, Apartado 644, E-48080 Bilbao, Spain}
\affiliation{IKERBASQUE, Basque Foundation for Science, Maria Diaz de Haro 3, 48013 Bilbao, Spain}
\affiliation{Department of Physics, Shanghai University, 200444 Shanghai, China}
\author{Yasser Omar}
\email{yasser.omar@lx.it.pt}
\affiliation{Instituto de Telecomunica\c{c}\~oes, Physics of Information and Quantum Technologies Group, Portugal}
\affiliation{Instituto Superior T\'ecnico, Universidade de Lisboa, Portugal}

\date{\today}

\begin{abstract}
Memristors, resistors with memory whose outputs depend on the history of their inputs, have been
used with success in neuromorphic architectures, particularly as synapses and non-volatile memories. However, to the best of our knowledge, no model for a network in which both the synapses and
the neurons are implemented using memristors has been proposed so far. In the present work we
introduce models for single and multilayer perceptrons based exclusively on memristors. We adapt
the delta rule to the memristor-based single-layer perceptron and the backpropagation algorithm
to the memristor-based multilayer perceptron. Our results show that both perform as expected
for perceptrons, including satisfying Minsky-Papert’s theorem. As a consequence of the Universal
Approximation Theorem, they also show that memristors are universal function approximators.
By using memristors for both the neurons and the synapses, our models pave the way for novel
memristor-based neural network architectures and algorithms. A neural network based on memristors could show advantages in terms of energy conservation and open up possibilities for other
learning systems to be adapted to a memristor-based paradigm, both in the classical and quantum
learning realms.
\end{abstract}

\maketitle


\section{Introduction}
\label{sec:introduction}
\begin{figure}[b!]
    \centering
    \includegraphics[width=0.8\columnwidth]{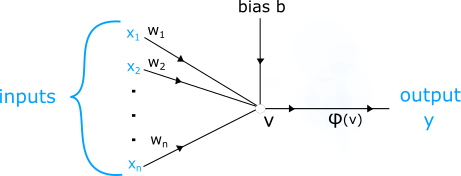}
    \caption{In a single-layer perceptron (SLP) the inputs ${x_i}$ are multiplied by their respective weights ${w_i}$ and added, together with a bias $b$ to form the net input to the SLP, $v$. The output $y$ of the SLP is given by some activation function, $\phi(v)$.}
    \label{fig:slp}
\end{figure}

\begin{figure}[htpb]
    \centering
    \includegraphics[width=0.8\columnwidth]{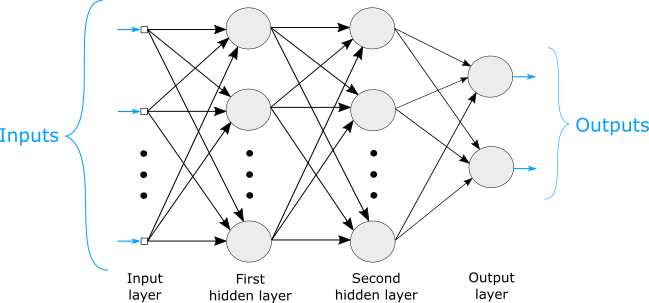}
    \caption{In a multilayer perceptron (MLP), single-layer perceptrons (SLP) are arranged in layers and connected to each other, with the outputs of the SLPs in the output layer being the outputs of the MLP. Here, each SLP is represented by a disc.}
    \label{fig:mlp}
\end{figure}

The perceptron, introduced by Rosenblatt in 1958 \cite{rosenblatt1958perceptron}, was one of the first models for supervised learning. In a perceptron, the inputs $x_1...x_n$ are linearly combined with coefficients given by the weights $w_1...w_n$, as well as with a bias $b$ to form the input $v$ to the neuron (see Fig. \ref{fig:slp}). $v$ is then fed into a non-linear function whose output is either $0$ or $1$. The goal of the perceptron is thus to find a set of weights $\{w_i\}$ that correctly assigns inputs $\{x_i\}$ to one of two predetermined binary classes. The correct weights for this task are found by an iterative training process, for instance the delta rule~\cite{widrow1960adaptive}. However, the perceptron is only capable of learning linearly separable patterns, as was shown in 1969 by Minksy and Papert \cite{minsky2017perceptrons}.
These limitations triggered a search for more capable models, which eventually resulted in the proposal of the multilayer perceptron. These objects can be seen as several layers of perceptrons connected to each other by synapses (see Fig. \ref{fig:mlp}). This structure ensures that the multilayer perceptron does not suffer from the same limitations as Rosenblatt's perceptron. In fact, the Universal Approximation Theorem \cite{cybenko1989approximation} states that a multilayer perceptron with at least one hidden layer of neurons and with conveniently chosen activation functions can approximate any continuous function to an arbitrary accuracy.

There are various methods to train a neural network such as the multilayer perceptron. One of the most widespread is the backpropagation algorithm, a generalization of the original delta rule~\cite{rumelhart1986learning}.

Artificial neural networks such as the multilayer perceptron have proven extremely useful in solving a wide variety of problems~\cite{rowley1998neural, devlin2014fast,ercal1994neural}, but they have thus far mostly been implemented in digital computers. This means that we are not profiting from some of the advantages that these networks could have over traditional computing paradigms, such as very low energy consumption and massive parallelization~\cite{jain1996artificial}. Keeping these advantages is, of course, of utmost interest, and this could be done if a physical neural network was used instead of a simulation on a digital computer. In order to construct such a network, a suitable building block must be found, with the memristor being a good candidate.

Besides these energetic considerations, exploring the fact that MLPs are universal function approximators our proposal of MLPs based only on memristors implies that memristive circuits can approximate any smooth function  $f:\mathbb{R}^n \rightarrow \mathbb{R}^m$ to arbitrary accuracy. 

The memristor was first introduced in 1971 as a two-terminal device that behaves as a resistor with memory \cite{chua1971memristor}. The three known elementary circuit elements, namely the resistor, the capacitor and the inductor, can be defined by the relation they establish between two of the four fundamental circuit variables: the current $i$, the voltage $u$, the charge $q$ and the flux-linkage $\phi$. There are six possible combinations of these four variables, five of which lead to widely-known relations: three from the circuit elements mentioned above, and two given by $q(t) = \int_{-\infty}^t i(\tau)d\tau$ and $\phi(t) = \int_{-\infty}^t u(\tau) d\tau$. This means that only the relation between $\phi$ and $q$ remains to be defined: the memristor provides this missing relation. Despite having been predicted in 1971 using this argument, it was not until 2008 that the existence of memristors was demonstrated at HP Labs \cite{strukov2008missing}, which led to a new boom in memristor-related research \cite{prodromakis2010review}. In particular, there have been proposals of how memristors could be used in Hebbian learning systems~\cite{soudry2013hebbian, cantley2011hebbian, he2014enabling},  in the simulation of fluid-like integro-differential equations \cite{barrios2018analog}, in the construction of digital quantum computers \cite{pershin2012neuromorphic} and of how they could be used to implement non-volatile memories \cite{ho2009nonvolatile}. 

The pinched current-voltage hysteresis loop inherent to memristors endows them with intrinsic memory capabilities, leading to the belief that they might be used as a building block in neural computing architectures \cite{traversa2015universal,pershin2010experimental,yang2013memristive}. Furthermore, the relatively small dimension of memristors, the fact that they can be laid out in a very dense manner and their non-volatile nature may lead to highly parallel, energy efficient neuromorphic hardware \cite{strachan2011measuring, jeong2016memristors, taha2013exploring, indiveri2013integration}. 

The possibility of using memristors as synapses in neural networks has been extensively studied. The wealth of proposals in this field can be broadly split into two groups: one related to spike-timing-dependent plasticity (STDP) and spiking neural networks (SNN) \cite{mostafa2015implementation, thomas2013memristor, ebong2012cmos, afifi2009implementation, querlioz2011simulation}, and the other to more traditional neural network models \cite{soudry2015memristor, hasan2014enabling, bayat2017memristor, negrov2017approximate, emelyanov2016first, wang2013memristive, yakopcic2013energy, demin2015hardware, duan2015memristor, prezioso2015training, wu2012synchronization, wen2018general, adhikari2012memristor}. The first group has a more biological focus, with its main goal being the reproduction of effects occurring in natural neural networks, rather than algorithmic improvements. In fact, the convergence of STDP-based learning is not guaranteed for general inputs \cite{soudry2015memristor}. The second group is more oriented towards neuromorphic computing and is composed of two major architectures, one based on memristor crossbars and another on memristor arrays. 

Despite all these results, and to the best of our knowledge, all existent proposals use memristors exclusively as synapses, with the networks' neurons being implemented by some other device. The main goal of this paper is thus to introduce a memristor-based perceptron, i.e., a single-layer perceptron (SLP) in which both synapses and neurons are built from memristors. It will be generalized to a memristor-based multilayer perceptron (MLP) and we will also introduce learning rules for both perceptrons, based on the delta rule for the SLP, and on the backpropagation algorithm for the MLP.

Recently the universality of memristors has been studied for Boolean functions~\cite{lehtonen2010two} and as a memcomputing equivalent of a Universal Turing Machine (Universal Memcomputing Machine~\cite{traversa2015universal}). However, to the best of our knowledge, it has not yet been shown that the memristor is a universal function approximator. This result will come as a consequence of the introduction of the above-mentioned memristor-based MLP.

\section{The memristor as a dynamical system}
In general, a current-controlled memristor is a dynamical system whose evolution is described by the following pair of equations~\cite{chua1971memristor}

\begin{subnumcases}{}
      V = R (\Vec{\gamma}, I) I \label{eq:gen_memristor}, \\
      \Dot{\Vec{\gamma}} = \Vec{f}(\Vec{\gamma},I) \label{eq:iv_update}.
\end{subnumcases}
The first one is Ohm's law and relates the voltage output of the memristor $V$ with the current input $I$ through the memristance $R (\Vec{\gamma}, I)$, which is a scalar function depending both on $I$ and on the set of the memristor's internal variables $\Vec{\gamma}$. This dependence of the memristance on the internal variables induces the memristor's output dependence on past inputs, i.e., this is the mechanism that endows the memristor with memory. The second equation describes the time-evolution of the memristor's internal variables by relating their time derivative, $\Dot{\Vec{\gamma}}$, to an $n$-dimensional vector function $\Vec{f}(\Vec{\gamma},I)$, depending on both previous values of the internal variables and the input of the memristor.

\subsection{Memristor-based Single-Layer Perceptron}
\begin{algorithm}[H]
  \caption{Delta rule for Single-layer Perceptron}
  \label{alg:SLP}
  \begin{algorithmic}
  \STATE \textbf{Initialization}
  \STATE Set the bias current $I_b$ to $0$.
  \STATE Initialize the weights $w_1$, $w_2$, $w_b$.
  \STATE Set the internal state variables $\gamma_1$, $\gamma_2$, $\gamma_3$ to $w_1$, $w_2$ and $w_b$, respectively. 
  \FOR{d in data}
      \HEADER{Forward Pass}
      \STATE Compute the net input to the perceptron: 
            \begin{equation}
                I = w_1x_1 + w_2x_2.
            \end{equation}
      \STATE Compute the perceptron's output:    
            \begin{equation}
                V =  g(I, \gamma_1, \gamma_2, \gamma_3).
            \end{equation}
      \ENDHEADER    
      \HEADER{Backward Pass}        
      \STATE Compute the difference $\Delta$ between the target output and the actual output:
        \begin{equation}
            \Delta = T - V.
        \end{equation}
      \STATE Compute the derivative of the activation function with respect to the net input, $g'$.

      \FOR{i in internal variables}
        \IF{$\Delta \geq 0$}
            \STATE Set the bias $I_b = I_{\gamma_i}$.
        \ELSE  
            \STATE Set the bias $I_b = -I_{\gamma_i}$.
        \ENDIF    
        \STATE Update $\gamma_i$ by inputting $I =  \Delta x_{i}g' + I_b$. 
      \ENDFOR
      \STATE Update the weights by setting them to the updated values of the internal state variables.
      \STATE Set the bias $I_b = 0$.
      \ENDHEADER
  \ENDFOR
\end{algorithmic}
\end{algorithm}

Our goal is to implement a perceptron and an adaptation of the delta rule to train it using only a memristor. To this end, we use the memristor's internal variables to store the SLP's weights and the learning rate. Equation~\eqref{eq:iv_update} allows us to control the evolution of the memristor's internal variables and implement a learning rule. If, for example, we want to implement a SLP with two inputs we need a memristor with four internal variables, two of them to store the weights of the connections between the inputs and the SLP, a third one to store the SLP's bias weight and another for the learning rate.

Let us then consider a memristor with four internal state variables, from now on labeled by $\Vec{\gamma} = (\gamma_1, \gamma_2,\gamma_3, \gamma_4)$ and in which $\Vec{f} = (f_1,f_2,f_3, f_4)$. It could be difficult to externally control multiple internal variables. However, a possible solution is to use several memristors with the chosen requirements and with an externally controlled internal variable each.



In order to understand the form of these functions, we must remember that we expect different behaviours from the perceptron depending on the stage of the algorithm. In the forward propagation stage, the weights must remain constant to obtain the output for a given input. In this phase the internal variables must not change. On the other hand, in the backpropagation stage, we want to update the perceptron's weights by changing the internal variables. However, it may happen that the update is different for each of the weights, so we need to be able to change only one of the internal variables without affecting the others. 

There are thus three different possible scenarios in the backpropagation stage: we want to update $\gamma_1$, while $\gamma_2$ and $\gamma_3$ should not change; we want to update $\gamma_2$, while $\gamma_1$ and $\gamma_3$ should not change, and we want to update $\gamma_3$, while $\gamma_1$ and $\gamma_2$ should not change.
To conciliate this with the fact that a memristor takes only one input, we propose the use of threshold-based functions, as well as a bias current $I_b$, for the evolution of the internal variables
\begin{align}
      V(t) &= g (I, \gamma_1, \gamma_2, \gamma_3), \label{eq:slp_memristor}\\
      \begin{split}
      \Dot{\gamma_i} &= (I-I_b) \left(\theta (I-I_{\gamma_i}) - \theta(I-\left(I_{\gamma_i} + a)\right)\right)\\
      &+ (I + I_b) \left(\theta(-I-I_{\gamma_i}) - \theta(-I-(I_{\gamma_i} + a))\right),          
      \end{split}
\label{eq:slp_gamma1}
\end{align}
where $g$ is an activation function, $\theta$ is the Heaviside function, $I_{\gamma_i}$ is the threshold for the internal variable $\gamma_i$ and $a$ is a parameter that determines the dimension of the threshold, i.e., the range of current values for which the internal variables are updated. The first term of the update function can only be non-zero if the input current is positive, whereas the second term can only be non-zero if the input current is negative, allowing us to both increase and decrease the values of the internal variables. If $I_{\gamma_1}$, $I_{\gamma_2}$ and $I_{\gamma_3}$ are sufficiently different from each other and from zero, we can reach the correct behaviour by choosing the memristor's input appropriately. The thresholds and the $a$ parameter are thus hyperparameters that must be calibrated for each problem. In the aforementioned construction in which our memristor with three internal variables is constructed as an equivalent memristor, we can also use an external current or voltage control to keep the internal variable fixed. In fact, this is how it is usually addressed experimentally \cite{yang2013memristive,xia2009memristor,yu2015dynamic,budhathoki2013composite}. Therefore, we can assume that this construction is possible. It is important to note that, in an experimental implementation, this threshold system does not need to be based on the input currents' intensities. It can, for instance, be based on the use of signals of different frequencies for each of the internal variables or in the codification of the signals meant for each of the internal variables in AC voltage signals. 

We are now ready to present a learning algorithm for our SLP based on the delta rule, which is described in Algorithm~\ref{alg:SLP}. In case one wants to generalize this procedure to an arbitrary number of inputs $n$, this can be trivially achieved by using a memristor with $n+1$ internal variables and adapting Algorithm~\ref{alg:SLP} accordingly. 

\subsection{Memristor-based Multilayer Perceptron}
\begin{algorithm}[H]
  \caption{Backpropagation for Multilayer Perceptron}
  \label{alg:MLP}
  \begin{algorithmic}
  \HEADER{Initialization}
  \STATE Set the bias current $I_b$ to $0$.
  \STATE Initialize the weights $\{w_{ij}\}$ and $\{w_{b_k}\}$.
  \STATE Set the internal variable $\gamma_{ij}$ of each connection memristor $ij$ to the respective connection weight $w_{ij}$.
  \STATE Set the internal variable $\gamma_{k}$ of each connection memristor $k$ to the respective bias weight $w_{b_k}$.
  \ENDHEADER
  \FOR{d in data}
    \HEADER{Forward Pass}
        \FOR{l in layers}
          \STATE Compute the output of each connection memristor $ij$ in layer $l$: 
                \begin{equation}
                    V_{ij} (w_{ij}, I) = w_{ij}I.
                \end{equation}
          \STATE Sum the outputs of the connection memristors connected to each node memristor $k$ in layer $l$
            \begin{equation}
                \text{in}_k = \sum I_{ik}
            \end{equation}
          \STATE Compute the node memristor's output:
            \begin{eqnarray*}
                V_k = R_{\text{OFF}}\left(1 - \frac{\gamma_{b_k}}{D}+\frac{R_{\text{ON}}}{R_{\text{OFF}}}\frac{\gamma_{b_k}}{D}\right)\text{in}_k.
            \end{eqnarray*}
            \normalsize
        \ENDFOR
    \ENDHEADER
    \HEADER{Backward Pass}
      \FOR{k in output layer}
      \STATE Compute the difference $\Delta$ between the target output and the actual output of the node memristor:
        \begin{equation}
            \Delta_k = T_k - V_k.
        \end{equation}
      \STATE Compute the local gradient of the node memristor using Equation~\eqref{eq:gradient_out}.
        \ENDFOR
     \FOR{layer in hidden layers}
        \FOR{node in layer}
            \STATE Compute the local gradient of node memristor $l$ in layer using Equation~\eqref{eq:gradient_hidden}.
        \ENDFOR
    \ENDFOR
    \FOR{connection in connections}
        \STATE Compute the weight update. 
        \STATE Set the bias current: $I_b = I_{\gamma_{ij}}$.
        \STATE Update the connection memristor's internal variable by inputting $I = \Delta w_{ij} + I_{b}$ to it.
        \STATE Update the connection's weight by setting it to the updated value of the respective internal variable.
    \ENDFOR    
    \FOR{node in nodes}
        \STATE Compute the bias weight update according to Equation~\eqref{eq:bias_update}.
        \STATE Set the bias current: $I_b = I_{\gamma_{b}}$.
        \STATE Update the node memristor's internal variable by inputting $I = \Delta w_{k} + I_{b}$.
        \STATE Update the bias weight by setting it to the updated value of the respective internal variable.
    \ENDFOR
    \ENDHEADER
  \ENDFOR
\end{algorithmic}
\end{algorithm}

In this model, memristors are used to emulate both the connections and the nodes of a MLP. In principle, the nodes could be emulated by non-linear resistors, but using memristors allows us to take advantage of their internal variable to implement a bias weight, which in some cases proves fundamental for a successful network training.

The equations describing the evolution of the memristor at each node in this model are the same as in the seminal HP Labs paper \cite{strukov2008missing}. We have chosen the experimentally tested set
\begin{align}
V(t) = \left(R_{\text{ON}}\frac{\gamma(t)}{D} + R_{\text{OFF}}\left(1 - \frac{\gamma(t)}{D}\right)\right) I(t), \label{eq:node_activation}\\
\Dot{\gamma} = \begin{cases}
\mu_V \frac{R_{\text{ON}}}{D}I(t) - I_{\gamma} & \text{if   } \mu_V \frac{R_{\text{ON}}}{D}I(t) > I_{\gamma}, \\
0 & \text{o.w.}
\end{cases} \label{eq:iv_evolution}
\end{align}
Here, $R_{\text{ON}}$ and $R_{\text{OFF}}$ are, respectively, the doped and undoped resistances of the memristor, $D$ and $\mu_V$ are physical memristor parameters, namely the thickness of its semiconductor film and its average ion mobility, and $I_{\gamma}$ is a threshold current playing the same role as the $I_{\Vec{\gamma}}$ in the model for the memristor-based SLP introduced above. Equation~\eqref{eq:node_activation} can be approximated by
\begin{equation}
    V(t) = R_{\text{OFF}} \left(1 - \frac{\gamma(t)}{D}\right) I(t),
\end{equation}
since we have that $\frac{R_{\text{ON}}}{R_{\text{OFF}}} \approx \frac{1}{100}$. If, for instance, we impose a constant current input $I$ to the memristor for a time $t$, the output is given by



\begin{equation}
    V(t) \propto -I^2t \label{eq:outputforconstantI}.
\end{equation}

It is then possible to implement non-linear activation functions starting from Equation~\eqref{eq:node_activation}, which is an important condition for the universality of neural networks \cite{hornik1991approximation}.

Looking now at synaptic memristors, their evolution is described by

\begin{align}
&V(t) = \gamma(t) I(t), \\
&\Dot{\gamma} = \left(\mu_V \frac{R_{\text{ON}}}{D} I(t) - I_{\gamma}\right)\theta\left(\mu_V \frac{R_{\text{ON}}}{D} I(t) - I_{\gamma}\right).
\end{align}

In synaptic memristors, the internal variable $\gamma$ is used to store the weight of the respective connection, whereas in node memristors the internal variable is used to store the node's bias weight.

As explained before, the node memristors are chosen to operate in a non-linear regime, which allows us to implement non-linear activation functions. On the other hand, we choose a linear regime for synaptic memristors, which allows us to emulate the multiplication of weights by signals.

It must be mentioned that Equation~\eqref{eq:iv_evolution} is only valid for $ \gamma \in [0,D]$. If we were to store the network weights in the internal variables using only a rescaling constant $A$, i.e., $w = A\gamma$, then the weights would all have the same sign. Although convergence of the standard backpropagation algorithm is still possible in this case~\cite{dickey1993optical}, it is usually slower and more difficult, so it is convenient to redefine the variable \cite{strukov2008missing} $D \rightarrow D'$ so that the interval of the internal variable in which Equation~\eqref{eq:iv_evolution} is valid becomes $[-D'/2,D'2]$. Using a rescaling constant $B$, the network weights can then be in the interval $[-BD'/2,BD'/2]$.

The new learning algorithm is an adaptation of the backpropagation algorithm, chosen due to its widespread use and robustness. In our case, the activation function of the neurons is the function that relates the output of a node memristor with its input, as seen in Equation~\eqref{eq:node_activation}. The local gradients of the output layer and hidden layer neurons are respectively given by:
\begin{numcases}{}
      \text{Output: }\delta_k = T_k \phi'\left(\sum_i V_{ik}\right) \label{eq:gradient_out},\\
      \text{Hidden: }\delta_k = \phi'\left(\sum_i V_{ik}\right)\sum_l\delta_lw_{kl}. \label{eq:gradient_hidden}
\end{numcases}
In Equation~\eqref{eq:gradient_out}, $T_k$ denotes the target output for neuron $k$ in the output layer. In Equations~\eqref{eq:gradient_out} and~\eqref{eq:gradient_hidden}, $\phi'$ is the derivative of the neuron's activation function with respect to the input to the neuron $\sum_i V_{ik}$. Finally, in Equation~\eqref{eq:gradient_hidden}, the sum $\sum_l\delta_lw_{kl}$ is taken over the gradients of all neurons $l$ in the layer to the right of the neuron that are connected to it by weights $w_{kl}$. The update to the bias weight of a node memristor is given by:
\begin{equation}
    \Delta w_k = \eta \delta_k,
    \label{eq:bias_update}
\end{equation}
where $\eta$ is the learning rate. The connection weight $w_{ij}$ is updated using $\Delta w_{ij} = \eta \delta_j V_i$, where $\delta_j$ is the local gradient of the neuron to the right of the connection, and $V_i$ is the output of the neuron to the left of the connection.


We count now with all necessary elements to adapt the backpropagation algorithm for our memristor-based MLP, as described in Algorithm \ref{alg:MLP}.


\section{Simulation results}
In order to test the validity of our SLP and MLP, we tested their performance on three logical gates: OR, AND and XOR. The first two are simple problems which should be successfully learnt by SLP and MLP, whereas only the MLP should be able to learn the XOR gate, due to Minsky-Papert's theorem. 

The Glorot weight initialization scheme \cite{glorot2010understanding} was used for all simulations, as it has been shown to bring faster convergence in some problems when compared to other initialization schemes. In this scheme the weights are initialized according to $\mathcal{U}(-1, 1)$, weighed by $\sqrt{\frac{6}{n_{in} + n_{out}}}$, where $n_{in}$ and $n_{out}$ are the number of neurons in the previous and following layers, respectively. The data sets used contain $100$ randomly generated labeled elements, which were shuffled for each epoch, and the cost function is:
\begin{equation}
E = \frac{1}{2}(T - O)^2,
\label{eq:cost}
\end{equation}
where $T$ is the target output and $O$ the actual output.
\subsection{Single-Layer Perceptron Simulation Results}
\begin{figure}[ht]
    \includegraphics[width=\columnwidth]{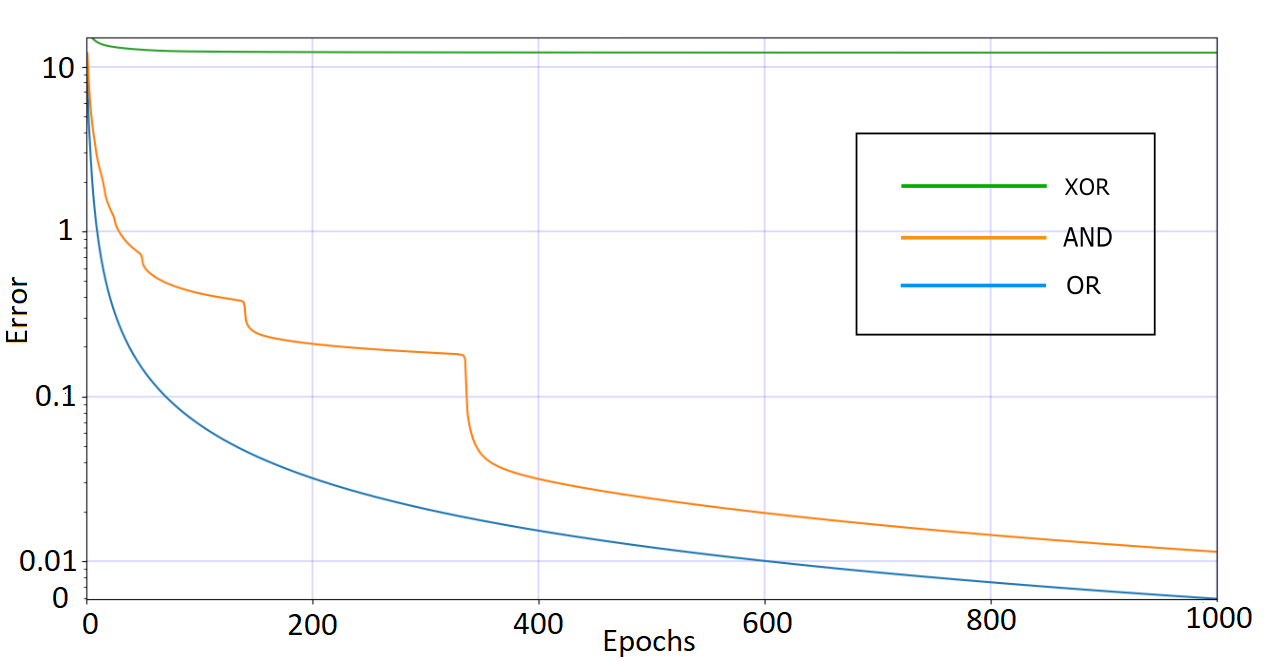}
    \caption{Evolution of the learning progress of our single-layer perceptron (SLP), quantified by its total error, given by Equation \eqref{eq:tot_error}, for the OR, AND and XOR gates over $1000$ epochs. The total error of our SLP for the OR and AND gates goes to $0$ very quickly, indicating that our SLP successfully learns these gates. The same is not true for the XOR gate, which our SLP is incapable of learning, in accordance with Minksy-Papert's theorem \cite{minsky2017perceptrons}.}
    \label{fig:SLP}
\end{figure}

For the SLP, a learning rate of $0.1$ was used for all tested gates, a value set by trial and error. The metric we used to evaluate the evolution of the network's performance on a given problem was its total error over an epoch, which is given by Equation \eqref{eq:tot_error}.
\begin{equation}
E_{\text{total}} = \sum_j E_j = \frac{1}{2}\sum_j (T_j - O_j)^2, \label{eq:tot_error}    
\end{equation}
where the sum is taken over all elements in the training set. In Fig.~\ref{fig:SLP}, the evolution of the total error over $1000$ epochs, averaged over $100$ different realizations of the starting weights, is plotted.\\

We observe that our SLP successfully learns the gates OR and AND, with the total error falling to $0$ within $200$ epochs, as expected from a SLP. However, the total error of our SLP for the XOR gate does not go to zero, which means that it is not able to learn this gate, in accordance with Minsky-Papert's theorem. 

\subsection{Multilayer Perceptron Simulation Results}
\begin{figure}[ht]
    \centering
    \includegraphics[width=\columnwidth]{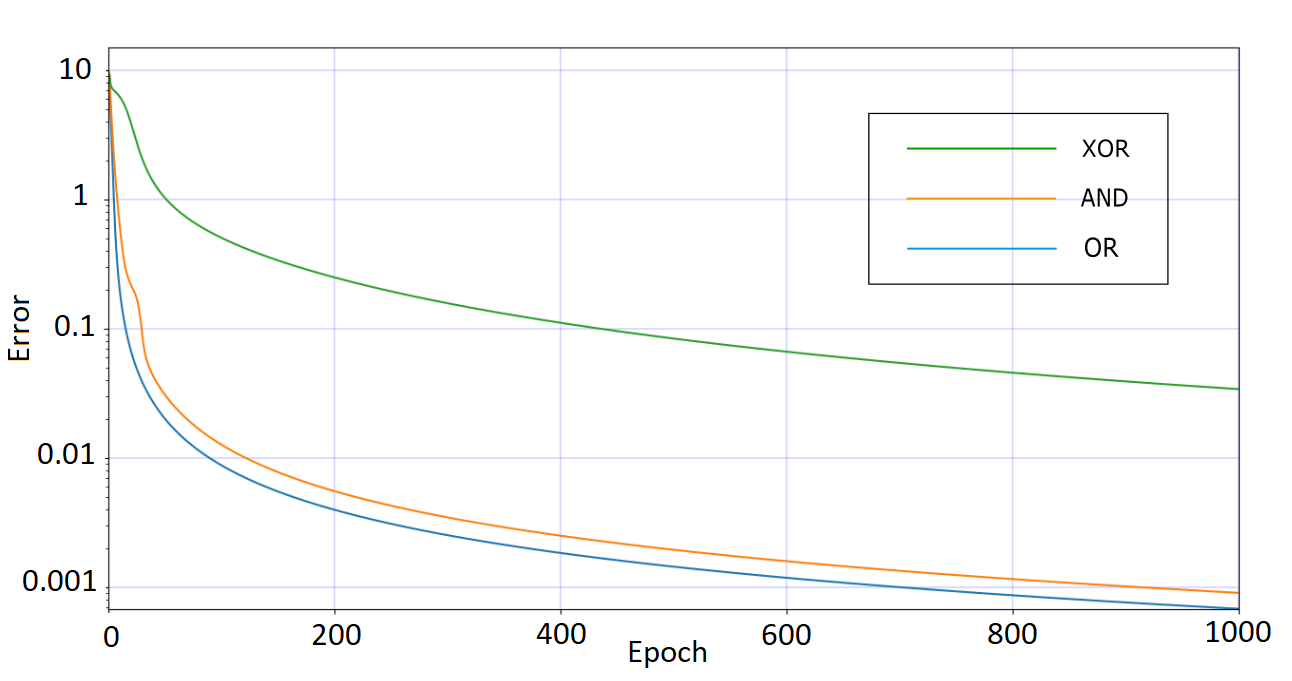}
 \caption{Evolution of the learning progress of our multilayer perceptron (MLP), quantified by its total error, given by Equation \eqref{eq:tot_error} for the OR, AND and XOR gates over $1000$ epochs. As can be seen, the total error of our MLP for the these gates approaches $0$, indicating that it successfully learns all three gates.}
  \label{fig:MLP}
\end{figure}

The structure of the network was chosen following~\cite{walczak1999heuristic}. There, a network with one hidden layer of two neurons is recommended for the case of two inputs and one output. As noted in~\cite{walczak1999heuristic}, networks with only one hidden layer are capable of approximating any function, although in some problems, adding extra hidden layers improves the performance. However, the results obtained by employing only one hidden layer are satisfactory, thus there is no need for a more complex network structure. There is also the matter of how many neurons must be employed in the hidden layer. In this case, there is a trade-off between speed of training and accuracy. A network with more neurons in the hidden layer counts with more free parameters, so it will be able to output a more accurate fit, but at the cost of a longer time required to train the network. A rule of thumb for choosing the number of neurons in the hidden layer is to start with an amount that is between the number of inputs and the number of outputs and adjust according to the results obtained. This leads to two neurons for the hidden layer and, similarly to what happened with the number of hidden layers, the results obtained using two neurons in the hidden layer are sufficiently accurate, so there was no need to try other structures. The learning rates used, which we have chosen through trial and error, are $0.1$ for the OR and AND gates, and $0.01$ for the XOR gate. In Fig.~\ref{fig:MLP}, the evolution of the total error over $1000$ epochs, averaged over $100$ different realizations of the starting weights, is plotted.\\

As was the case for our SLP, our MLP successfully learns the OR and AND gates. In fact, it is able to learn them faster than our SLP, which is a consequence of the larger number of free parameters. Additionally, it is able to learn the XOR gate, indicating that it behaves as well as a regular MLP.

In summary, both memristor-based perceptrons behave as expected. Our SLP is able to learn the OR and AND gates, but not the XOR gate, so it is limited to solving linearly separable problems, just as any other single-layer neural network. However, our MLP is not subject to such a limitation and it is able to learn all three gates.

\subsection{Receiver Operating Characteristic Curves}
\begin{figure}[H]
    \centering
    \includegraphics[width=\columnwidth]{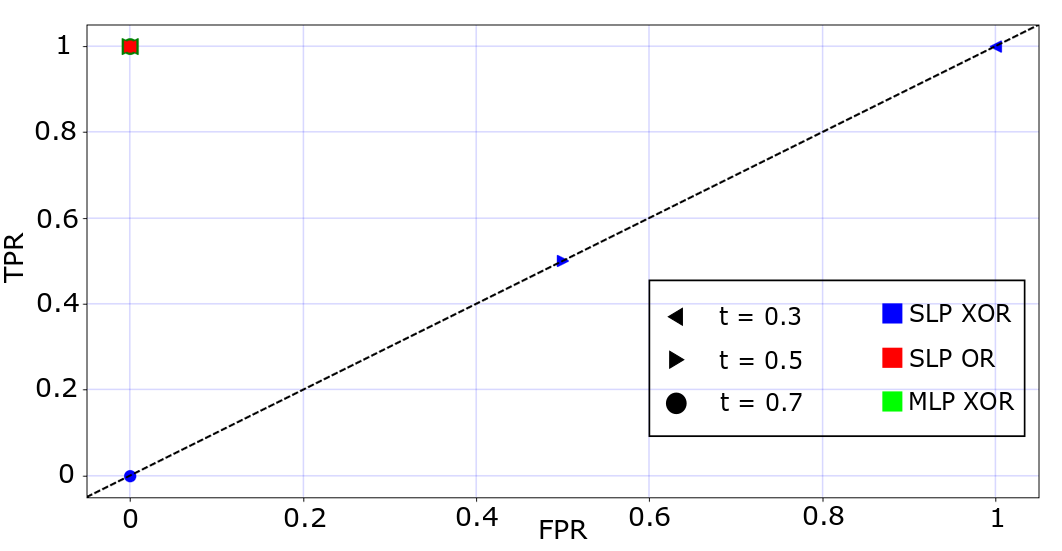}
 \caption{ROC curves obtained with the SLP for the OR and XOR gates, and with the MLP for the XOR gate. The thresholds used were $t = 0.3$, $0.5$ and $0.7$ We can see that the SLP correctly classifies the inputs for the OR gate every time, but it does not perform better than random guessing for the XOR gate, as expected. On the other hand, the MLP correctly classifies the XOR gate inputs every time.}
  \label{fig:ROC_ba}
\end{figure}

As another measure of the perceptrons' performance, we show in Fig.~\ref{fig:ROC_ba} the receiver operating characteristic (ROC) curves obtained with perceptrons trained for $500$ epochs on data sets of size $100$. The curves shown were obtained using a SLP trained for the OR gate, a SLP trained for the XOR gate and a MLP trained for the XOR gate, with thresholds of $t = 0.3$, $0.5$ and $0.7$ for each. Again, we see that the SLP is capable of learning the OR gate but not XOR, since it correctly classifies the inputs for OR every time, but its performance is equivalent to random guessing for XOR. We can also see that the MLP is capable of learning the XOR gate, since it correctly classifies its inputs every time. The learning rates used in training were $0.1$ for the SLP on both gates and $0.01$ for the MLP on XOR gate, as explained in the previous subsection.

\section{Conclusion}

In this paper, we introduced models for single and multilayer perceptrons based exclusively on memristors. We provided learning algorithms for both, based on the delta rule and on the backpropagation algorithm, respectively. Using a threshold-based system, our models are able to use the internal variables of memristors to store and update the perceptron's weights. We also ran simulations of both models, which revealed that they behaved as expected, and in accordance with Minsky-Papert's theorem. Our memristor-based perceptrons have the same capabilities of regular perceptrons, thus showing the feasibility and power of a neural network based exclusively on memristors.

To the best of our knowledge, our neural networks are the first ones in which memristors are used as both the neurons and the synapses. Due to the Universal Approximation Theorem for multilayer perceptrons, this implies that memristors are universal function approximators, i.e., they can approximate any smooth function  $f:\mathbb{R}^n \rightarrow \mathbb{R}^m$ to arbitrary accuracy, which is a novel result in their characterization as devices for computation.

Our models also pave the way for novel neural network architectures and algorithms based on memristors. As previously discussed, such networks could show advantages in terms of energy optimization, allow for higher synaptic densities and open up possibilities for other learning systems to be adapted to a memristor-based paradigm, both in the classical and quantum learning realms. In particular, it would be interesting to try to extend these models to the quantum computing paradigm, using a recently proposed quantum memristor \cite{pfeiffer2016quantum}, and its implementation in  quantum technologies, such as superconducting circuits \cite{salmilehto2017quantum} or quantum photonics \cite{sanz2017quantum}.

\begin{acknowledgments}
Work by FS was supported in part by a New Talents in Quantum Technologies scholarship from the Calouste Gulbenkian Foundation. FS and YO thank the support from Funda\c{c}\~{a}o para a Ci\^{e}ncia e a Tecnologia (Portugal), namely through programme POCH and projects UID/EEA/50008/2013 and IT/QuNet, as well as from the project TheBlinQC supported by the EU H2020 QuantERA ERA-NET Cofund in Quantum Technologies and by FCT (QuantERA/0001/2017), from the JTF project NQuN (ID 60478), and from the EU H2020 Quantum Flagship projects QIA (820445) and QMiCS (820505). MS and ES are grateful for the funding of Spanish MINECO/FEDER FIS2015-69983-P and Basque Government IT986-16. This material is also based upon work supported by the U.S. Department of Energy, Office of Science, Office of Advance Scientific Computing Research (ASCR), under field work proposal number ERKJ335. 
\end{acknowledgments}

\bibliographystyle{IEEEtran}
\bibliography{IEEEabrv,biblio.bib}

\end{document}